\documentclass[lettersize,journal]{IEEEtran}
\usepackage{amsmath,amsfonts}
\usepackage{algorithmic}
\usepackage{algorithm}
\usepackage{array}

\usepackage{subcaption}   
\usepackage{url}
\usepackage{textcomp}
\usepackage{stfloats}
\usepackage{url}
\usepackage{verbatim}
\usepackage{graphicx}
\usepackage{cite}
\usepackage{booktabs, pifont, makecell}
\usepackage{multirow}   
\usepackage{threeparttable}   
\usepackage{caption}   
\usepackage{xcolor}

\usepackage{amsmath,amsthm,amstext,amsfonts,amssymb,mathrsfs}
\theoremstyle{definition}
\newtheorem{definition}{Definition}

\newtheorem{assumption}{Assumption}

\newtheorem{remark}{Remark}

\begin{document}

\title{Adaptive Electromagnetic Transient-level Digital Twins for Data Centers via In-Context Learning with LLMs}

\author{
  Amit~Jena,~\IEEEmembership{}
  Bin~Wang,~\IEEEmembership{Senior Member,~IEEE,}
  Aoxia~Chen,
  Brad~Chadwell,
  Billy~Yancey,
  and~Le~Xie,~\IEEEmembership{Fellow,~IEEE}
  \thanks{A. Jena is with the Department of Electrical and 
  Computer Engineering, Texas A\&M University, College Station, 
  TX 77843 USA (e-mail: amit\_jena@tamu.edu). \textit{(Corresponding author: Amit Jena.)}}
  \thanks{B. Wang is with Advanced Technology Solutions, ISO 
  New England, Holyoke, MA 01040 USA (e-mail: bwang@iso-ne.com).}
  \thanks{A. Chen, B. Chadwell, and B. Yancey are with Electric 
  Power Engineers (EPE), Austin, TX USA (e-mail: 
  kchen@epeconsulting.com; bchadwell@epeconsulting.com; 
  byancey@epeconsulting.com).}
  \thanks{L. Xie is with the School of Engineering and Applied 
  Sciences, Harvard University, Boston, MA 02134 USA 
  (e-mail: xie@seas.harvard.edu).}
}


\maketitle

\begin{abstract}
The rapid growth of artificial intelligence (AI) workloads has significantly increased data center (DC) power demand. Frequent operational and configurational changes, such as load surges caused by dynamic workload allocation, can induce electromagnetic transients (EMTs) whose accurate prediction is critical for reliable DC operation and maintenance. Although high-fidelity EMT solvers are widely adopted for transient analysis, their computational cost limits large-scale and multi-scenario studies. Digital twins address this computational bottleneck by emulating the behavior of a physical system and enabling fast prediction in place of expensive high-fidelity simulation. However,
data-driven digital twins are non-adaptive and fail to generalize once the DC configuration changes. To address this limitation, this paper proposes an adaptive EMT-level digital twin for DCs based on in-context learning (ICL) with large language models (LLMs). We formulate the EMT prediction task as a system identification problem under parametric uncertainty and develop a GPT-2-based framework that adapts to unseen DC configurations at inference time from a small set of EMT trajectory samples, without model retraining.  The proposed framework is validated on three representative EMT scenarios and benchmarked against three baseline non-adaptive twins. Results demonstrate superior prediction accuracy, improved adaptability to unseen operating conditions, and substantial computational speedups compared to high-fidelity EMT simulation.
\end{abstract}


\section{Introduction}
\IEEEPARstart{D}{ata} center (DC) energy demand has grown by 12\% since 2017 \cite{iea2025energyai}, now representing 1.5\% of global electricity consumption \cite{iea2025energyai}. The increasing usage of large language models (LLMs) and other compute-intensive AI tools has significantly contributed to this surge in demand. The rapid expansion of hyperscale DC facilities further amplifies the net energy consumption and puts additional stress on the power grid \cite{iea2025surge}. Given the heavy dependence of the technology and IT infrastructure on DCs, ensuring reliable power and energy management of these critical facilities has become a task of paramount importance.

Most DCs consist of power-electronic-heavy components such as uninterruptible power supplies (UPS) \cite{li2026ups}, variable frequency drives (VFDs) \cite{ai2023vfd}, and inverter-based resources (IBRs) \cite{seshmasetti2025review}. These components can introduce fast electromagnetic transients (EMTs) under abrupt changes in DC operating conditions \cite{aidc2025grid}. For instance, sudden load surges caused by unexpected changes in IT workloads \cite{aidc2025grid} may lead to significant EMT behavior. Similarly, fast switching between grid-connected and islanded (off-grid) operating modes may result in sub-cycle voltage and current oscillations in a DC \cite{azizi2026strengthening}.
Such EMT phenomena are important to analyze and predict, as they may otherwise lead to severe consequences, including failures of critical DC components, DC-wide power outages, and, at sufficient scale, adverse impacts on the upstream transmission system. Traditional simulation tools like PSCAD, EMTP-RV, and PLECS provide high-fidelity EMT modeling with microsecond-level time resolution. However, full EMT runs across large time horizons remain computationally intensive and slow \cite{guest2023speeding}, often requiring high-end computing infrastructure and several hours of wall-clock time for a single scenario. These computational bottlenecks make large-scale what-if analysis and planning impractical, thereby limiting efficient maintenance, reliability assessment, and failure prognosis in modern DCs.

Data-driven digital twins have emerged as promising alternatives to computationally expensive power system solvers (including EMT solvers) by providing fast and lightweight surrogates suitable for real-world deployment \cite{ogiesoba2026real}. A rich body of existing literature has explored classical digital twins based on optimization algorithms \cite{liu2026digital} or control-theoretic formulations \cite{pandey2026state}. However, such model-based approaches often oversimplify the dynamics, which may result in degradation in prediction and control performance \cite{wu2025overview}. To address these challenges, recent studies have proposed data-driven methods that offer a richer hypothesis class to capture complex dynamical systems. Advanced machine learning  (ML) architectures like long short-term memory (LSTM) networks \cite{wu2025overview} and gradient boosting machines \cite{ogiesoba2026real} have been proposed as digital twins to emulate EMT simulators. These approaches provide orders-of-magnitude faster execution compared to conventional EMT solvers while maintaining high simulation fidelity and accuracy \cite{ogiesoba2026real}.

In the context of DCs and other large-scale electrical infrastructures, such digital twins have been proposed for improved thermal management \cite{jadhav2026digital}, power flow monitoring \cite{sado2025digital}, and efficient workload scheduling \cite{nicolae2026opendt}. Despite their strong modeling capability, these digital twins suffer 
from a major limitation: \textit{a lack of adaptability to new operating conditions and configurational changes}. This limitation significantly inhibits the deployment of such static digital twins in real-world DCs, which are prone to operational uncertainties and evolving hardware configurations. Existing digital twins often act like fixed-configuration emulators and become impractical when system configurations evolve over time. A straightforward remedy to train a separate digital twin for each DC configuration is computationally prohibitive and demands substantial storage resources. This brittleness under distribution shift motivates the need for an \textit{EMT-level adaptive digital twin} that can generalize across a broad range of DC configurations through efficient inference-time adaptation.

 This paper proposes an in-context learning (ICL)-based adaptive digital twin for predicting EMT-level voltage and current profiles in DCs. Specifically, the proposed model predicts the transient responses at the point of common coupling (PCC) under previously unseen DC configurations and operating conditions. Originally developed for natural language processing \cite{brown2020gpt3}, ICL enables a pre-trained LLM to adapt to new tasks at inference time using only a small set of input-output examples, without gradient-based retraining. Leveraging this capability, the proposed adaptive twin generalizes EMT prediction across varying DC configurations without parameter updates.

\begin{figure}[t]
    \centering
    \includegraphics[width=\linewidth]{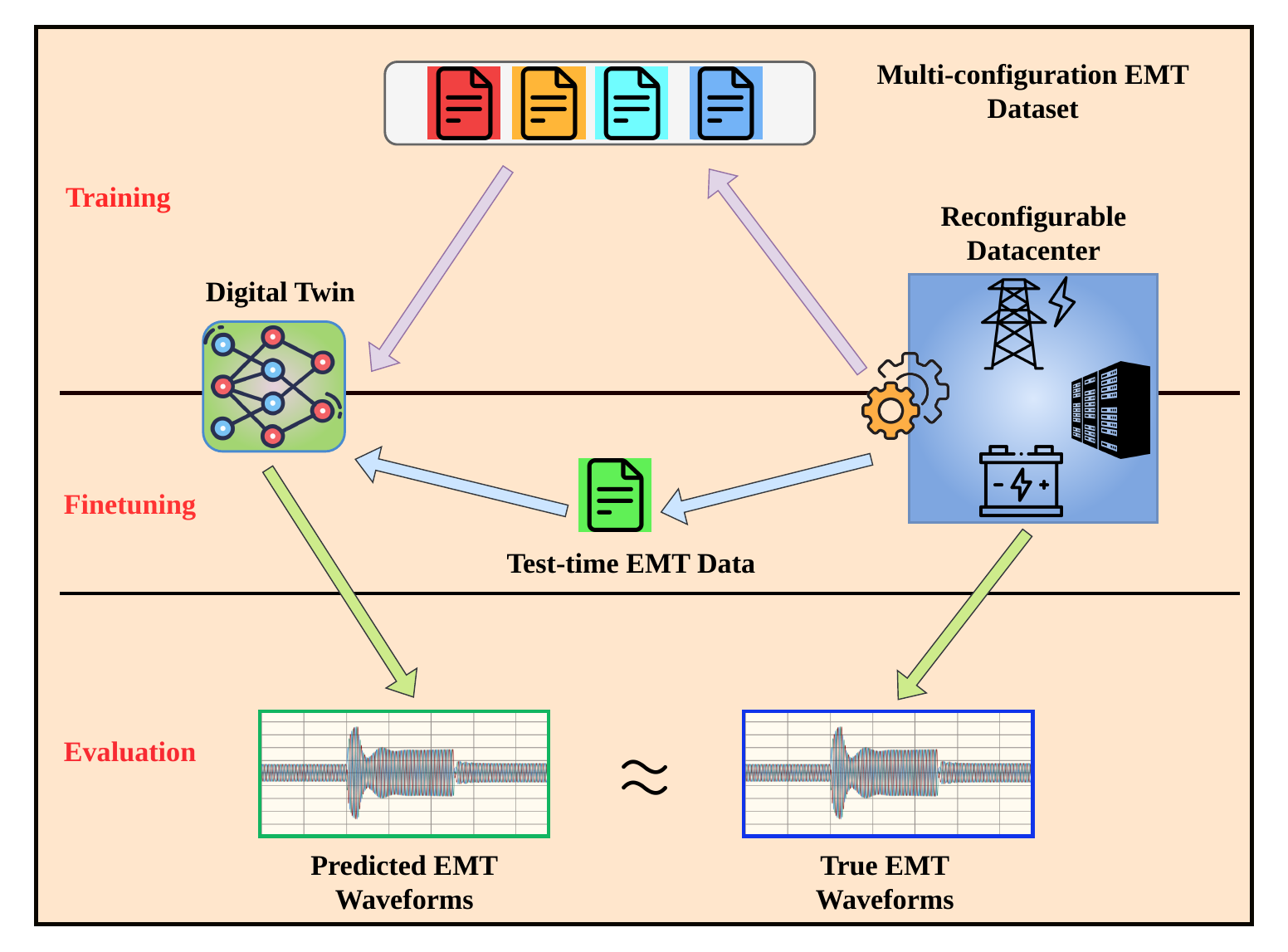}
    \caption{\textit{Overview of the Proposed ICL-based Adaptive Digital Twin Framework}. The twin is meta-trained on multi-configuration EMT data, adapted to unseen configurations using a few test-time samples, and evaluated by predicting EMT waveforms with high fidelity.}
    \label{fig:AD_twin}
\end{figure}

 To this end, we employ a GPT-2-based architecture trained on a multi-configuration EMT dataset generated from a high-fidelity DC EMT model. Key DC parameters, including power factors, load surge magnitudes, and UPS inverter ratings, are modeled as stochastic variables to construct a diverse training dataset. Treating abrupt changes in DC configurations or operating conditions as ``disturbances'', the adaptive twin generalizes to unseen scenarios by conditioning on a short post-disturbance EMT trajectory. A schematic overview of the proposed framework is shown in Fig. \ref{fig:AD_twin}. The main contributions of this paper are as follows:

 \begin{itemize}
     \item We formulate the adaptive EMT-level digital twin problem as a system identification task under parametric uncertainty, and propose a novel ICL-based solution that achieves inference-time adaptation across a broad range of data center configurations. The proposed model further achieves zero-shot adaptation using only a small set of in-context EMT trajectory samples. 
     \item We investigate three representative EMT transient instances arising from step-change in load, gradual-change in load, and switching between grid-connected and off-grid operating modes.
     \item We benchmark against three non-adaptive baselines, a classical ARMAX model, a long short-term memory (LSTM) network, and a physics-informed neural network (PINN) model, and demonstrate the superior predictive performance of the proposed ICL-based adaptive digital twin.
 \end{itemize}
 The remainder of this paper is organized as follows. Section II presents a generic data center EMT-level system model that captures all considered test cases. Section III formulates the 
 adaptive digital twin problem as a system identification task. Section IV introduces the proposed ICL-based adaptive digital twin. Section V presents case studies and simulation results. Section VI concludes the paper and outlines future research directions.
\section{Preliminaries and System Modeling}
\begin{figure*}[t]
    \centering
    \includegraphics[width=\linewidth]{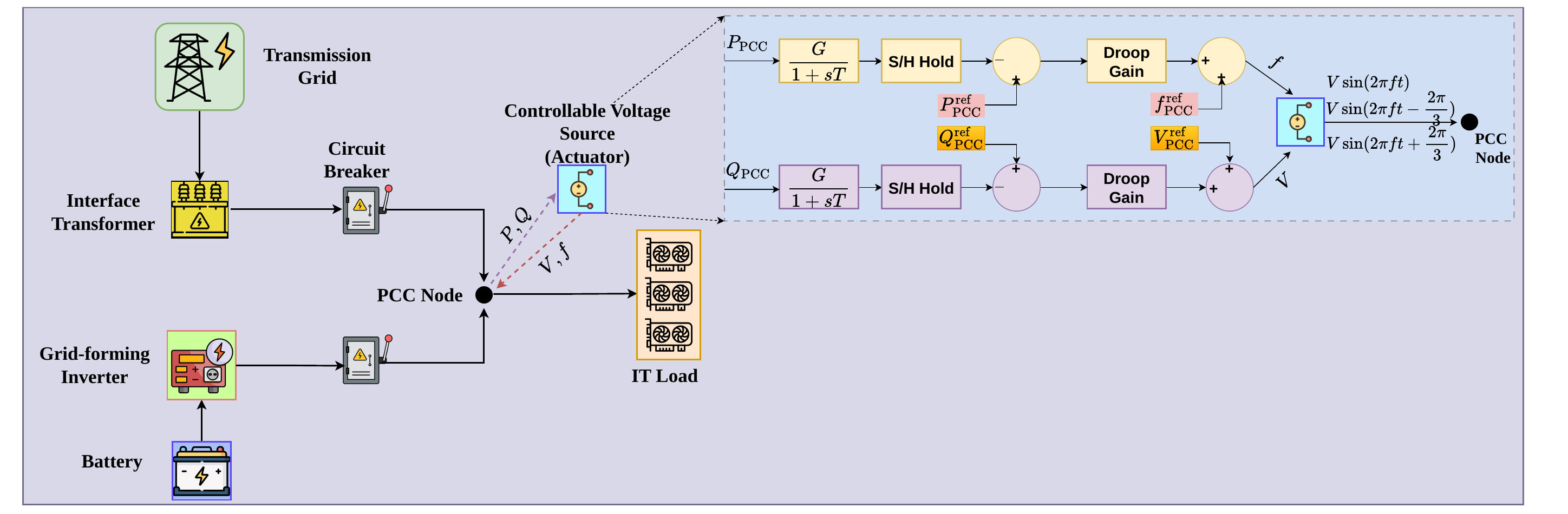}
    \caption{\textit{Schematic Diagram of the DC Power System Model.} The DC is supplied power through two different paths, thus leading to two distinct operating modes. (1) \textit{Grid-connected mode}: an upstream transmission grid and an interface transformer power the IT load, gated by a circuit breaker. (2) \textit{Off-grid mode}: A battery-backed grid-forming inverter provides power to the IT load, gated by a separate circuit breaker. The two paths join at the point of common coupling (PCC), which employs a droop-controlled actuator (inset) that regulates the PCC voltage by tracking active and reactive power ($P_\text{PCC}$, 
    $Q_\text{PCC}$) through low pass filtering, sample-and-hold, and $P-f$ and $Q-V$ droop control loops to set an internal three-phase voltage behind an output impedance.}
    \label{fig:DC_model}
\end{figure*}
This section presents a DC system architecture and develops a compact mathematical model for its EMT dynamics. We first introduce the physical configuration of the DC system, including its individual components and operating modes. The EMT dynamics is then represented as a parametric nonlinear differential-algebraic equation (DAE) system. Finally, we characterize three representative EMT scenarios, namely step-change-in-load-induced EMT, gradual-change-in-load-induced EMT, 
and mode-switching-induced EMT, which motivates the adaptive digital twin framework developed in subsequent sections. 
\subsection{DC EMT Model as a Parametric Nonlinear DAE}
\label{sec:DAE}
We consider a DC model as described in Fig. \ref{fig:DC_model}. It comprises an upstream transmission grid, an interface transformer, a battery, a grid-forming inverter, circuit breakers, a droop-controller actuator, and an IT load block. The full data center EMT model is represented as a  coupled nonlinear 
differential-algebraic equation (DAE) system of the form:
\begin{subequations}
\label{eq:gen_dyn}
    \begin{align}
\dot{\mathbf{x}}(t) &= \mathbf{f}(\mathbf{x}(t), \mathbf{y}(t), \mathbf{u}(t); \theta),\label{eq:gen_dyn_1} \\
\mathbf{0}          &= \mathbf{g}(\mathbf{x}(t), \mathbf{y}(t), \mathbf{u}(t); \theta), \label{eq:gen_dyn_2} \\
\mathbf{z}(t)       &= \mathbf{h}(\mathbf{x}(t), \mathbf{y}(t); \theta), \label{eq:gen_dyn_3}
\end{align}
\end{subequations}
where $\mathbf{x}(t)$ denotes the dynamic states, 
$\mathbf{y}(t)$ the algebraic states, $\mathbf{u}(t)$ the 
exogenous inputs, $\boldsymbol{\theta}$ the system parameters, 
and $\mathbf{z}(t) = [V_{abc}^\text{PCC}(t), I_{abc}^\text{PCC}(t)]$ 
the EMT outputs at the PCC, which are also the variables of interest. In what follows, we describe each DC component in detail.
\begin{itemize}
    \item \textbf{Upstream Transmission Grid}: The standard IEEE 39-bus system \cite{pscad_ieee39} is chosen as the upstream transmission network. The general rotor dynamics, exciter, and turbine-governor states are captured by $\mathbf{x}$, and the network power flow constraints partly constitute $\mathbf{g}(\cdot)$.
   \item \textbf{Interface Transformer}: It is modeled as a three-phase transformer characterized by its leakage reactance and winding resistance, forming algebraic constraints that contribute to $\mathbf{g}(\cdot)$.
    \item \textbf{Grid-forming Inverter and Battery}: The battery is modeled as an ideal DC source. The inverter dynamics include inner-loop current control and LC filter dynamics. The inverter variables, such as filter currents, capacitor voltages, voltage and current controller states, partly form $\mathbf{x}(t)$.
    \item \textbf{Droop-Controlled Actuator}: The actuator implements P-f and Q-V droop control as illustrated in Fig. \ref{fig:DC_model}. The entailed dynamics are:
    \begin{subequations}
    \begin{align}
    \dot{P}_\text{PCC}^{f}(t) &= -\tfrac{1}{T}P_\text{PCC}^{f}(t) + \tfrac{G}{T}P_\text{PCC}(t), \\
    \dot{Q}_\text{PCC}^{f}(t) &= -\tfrac{1}{T}Q_\text{PCC}^{f}(t) + \tfrac{G}{T}Q_\text{PCC}(t),
    \end{align}
    \end{subequations}
    where $P_\text{PCC}^f$ and $Q_\text{PCC}^f$ are the filtered active and reactive power measurements, $G$ is the measurement gain and $T$ is the low-pass filter 
    time constant. The droop outputs are described by:
    \begin{subequations}
    \label{eq:droop}
    \begin{align}
    f(t) &= f_\text{PCC}^\text{ref} - K_p (P_\text{PCC}^{f}(t) - P_\text{PCC}^\text{ref}), \\
    V(t) &= V_\text{PCC}^\text{ref} - K_q (Q_\text{PCC}^{f}(t) - Q_\text{PCC}^\text{ref}),
    \end{align}
    \end{subequations}
    where $K_p$ and $K_q$ are the active power-frequency and reactive power-voltage droop gains, respectively, and $\{P_\text{PCC}^\text{ref}, Q_\text{PCC}^\text{ref}, f_\text{PCC}^\text{ref}, V_\text{PCC}^\text{ref}\}$ denote the reference setpoints for active power, reactive power, frequency, and voltage magnitude at the PCC.
    Finally, the actuator produces the three-phase voltage $V \sin(2\pi f t + \phi_k)$, $\phi_k \in \{0, -2\pi/3, +2\pi/3\}$.
    \item \textbf{IT Load}: The aggregate IT load is modeled as a constant PQ load with time-dependent setpoints, $\{P_L(t), Q_L(t)\}$, serving as the primary exogenous input $\mathbf{u}(t)$. This aggregated representation implicitly incorporates the cascaded power conversion stages internal to the data center, including \textit{AC}/\textit{AC} isolation transformers, \textit{AC}/\textit{DC} rectifiers, and downstream \textit{DC}/\textit{DC} point-of-load converters that power the computing hardware. The internal converter dynamics operate at timescales separable from the PCC-level EMT profile, and are therefore omitted from the considered model.
    \item \textbf{PCC Node}: Kirchhoff's current law at the PCC leads to the following coupling algebraic constraint:
        \begin{equation}
        I_\text{grid}(t) + I_\text{inv}(t) + I_\text{act}(t) - I_\text{load}(t) = \mathbf{0},
    \end{equation}
    where $I_\text{grid}(t)$ is the three-phase current injected from the transmission grid through the interface transformer, $I_\text{inv}(t)$ is the current flowing from the grid-forming inverter branch, $I_\text{act}(t)$ is the current because of the droop-controlled actuator, and $I_\text{load}(t)$ is the current drawn by the IT load. The actuator current is determined by:
    \begin{equation}
    I_\text{act}(t) = \frac{V_\text{act}(t) - V^\text{PCC}(t)}{Z_\text{act}}
    \end{equation}
    where $V_{act}(t)$ is the three-phase voltage produced by the droop controllers described in \eqref{eq:droop} and $Z_{act}$ is the actuator's output impedance.
    \item \textbf{Circuit Breakers}: The circuit breaker states $s^\text{CB}_\text{grid}, s^\text{CB}_\text{inv} \in \{0, 1\}$ 
determine the data center operating mode:
\begin{align}
    \text{Grid-connected Mode}:&\quad s^\text{CB}_\text{grid} = 1, \; s^\text{CB}_\text{inv} = 0 \nonumber\\
    \text{Islanded Mode}:&\quad s^\text{CB}_\text{grid} = 0, \; s^\text{CB}_\text{inv} = 1 \nonumber
\end{align}
We note that the two breakers function in a mutually exclusive fashion. Specifically, simultaneous closure of both breakers is prohibited to avoid uncontrolled paralleling of the grid-forming inverter with the upstream grid, while simultaneous opening is not allowed to ensure a continuous power supply to the IT load. Mathematically,   $s^\text{CB}_\text{grid} + s^\text{CB}_\text{inv} = 1$  is ensured at all times.
\end{itemize}

\begin{assumption}[Ideal Battery Storage]
\label{assum:infinite_dc}
    The battery storage unit, which is connected to the grid-forming inverter, is assumed to have infinite energy capacity. Equivalently, the battery state of charge remains constant during the entire simulation horizon. Therefore, the battery recharging dynamics are ignored. This assumption isolates the EMT-level transient behavior at the PCC from longer-timescale energy management considerations, which are outside the scope of this work.  
\end{assumption}

\begin{assumption}[Strong Upstream Grid]
The upstream transmission grid is assumed to be strong at the high-voltage side of the interface transformer, maintaining a stiff three-phase voltage throughout the EMT events considered in this work. Consequently, the grid-side Th\'evenin impedance is small relative to the DC-internal impedances, and the PCC transient response is governed primarily by the DC components rather than by upstream grid dynamics. Therefore, wide-area transmission system phenomena are neglected and considered outside the scope of this study.
\end{assumption}

 In the following subsection, we describe three representative EMT scenarios and provide their corresponding mathematical formulations. 
\subsection{EMT Scenarios of Interest}
\label{sec:EMT_cases}
We consider three representative EMT scenarios that commonly occur in real-world DC operations. Specifically, these scenarios involve voltage and current transients at the PCC observed in the EMT domain and are caused by sudden operational changes.
Each scenario introduces a unique class of disturbance to the DAE system \eqref{eq:gen_dyn},  producing a characteristic EMT response.
\subsubsection{Step-Change-in-Load-Induced EMT}
The IT load experiences an abrupt transition between two operating points at time $t_s$ as:
\begin{equation}
\big(P_L(t), Q_L(t)\big) = 
\begin{cases}
\big(P_L^{(1)}, Q_L^{(1)}\big) & t < t_s \\
\big(P_L^{(2)}, Q_L^{(2)}\big) & t \geq t_s
\end{cases}
\end{equation}
These transitions arise from sudden workload bursts, such as simultaneous activation of large compute clusters. 
\subsubsection{Gradual-Change-in-Load-Induced EMT}
The IT load undergoes a smooth ramping between two operating points over a finite interval $[t_s, t_s + \Delta t]$ following a quadratic profile: 
\begin{equation}
\label{eq:gradual_load}
\begin{aligned}
&\big(P_L(t),Q_L(t)\big)\\
= &
\begin{cases}
\big(P_L^{(1)}, Q_L^{(1)}\big) & t < t_s \\
\big(P_L^{(1)} + \alpha^2(t)\Delta P, \; Q_L^{(1)} + \alpha^2(t)\Delta Q\big) & t_s \leq t \leq t_s + \Delta t \\
\big(P_L^{(2)}, Q_L^{(2)}\big) & t > t_s + \Delta t
\end{cases}
\end{aligned}
\end{equation}
where $\alpha(t) = (t - t_s)/\Delta t \in [0, 1]$, 
$\Delta P = P_L^{(2)} - P_L^{(1)}$, and 
$\Delta Q = Q_L^{(2)} - Q_L^{(1)}$. The quadratic profile captures workload escalations that accelerate over time, such as the progressive activation of compute clusters with warm-up periods.
\subsubsection{Mode-Switching-Induced EMT}
The data center transitions between grid-connected and 
islanded modes at switching instant $t_s$ via a toggle of the breaker states described by:
\begin{equation}
\left(s^\text{CB}_\text{grid}(t_s^+), s^\text{CB}_\text{inv}(t_s^+)\right) 
= \left(s^\text{CB}_\text{inv}(t_s^-), s^\text{CB}_\text{grid}(t_s^-)\right)
\end{equation}
where $t_s^-$ and $t_s^+$ denote the instants immediately before and after the switching event. The mutual exclusivity is preserved across this transition to ensure a constant power supply to the IT load. The switching event induces a change in current paths, resulting in EMT behavior at the PCC.

Across all three scenarios, the EMT responses at the PCC depend both on the disturbance characteristics and the underlying parametric configurations, making the resulting EMT profile sensitive to their variability.
\section{Problem Formulation}
\label{sec:prob_form}
We assume that the algebraic constraint \eqref{eq:gen_dyn_2} is locally solvable, so that $\mathbf{y}(t)$ could be expressed as a function of $\mathbf{x}(t)$, $\mathbf{u}(t)$, and $\theta$. More concretely, the Jacobian $\partial \mathbf{g}/\partial \mathbf{y}$ is assumed to be non-singular in a neighborhood of the operating trajectory, which leads to the satisfaction of the implicit function theorem\cite{rudin2021principles} and guarantees the following map: 
\begin{equation}
\label{eq:y_sep}
\mathbf{y}(t) = \tilde{\mathbf{g}}(\mathbf{x}(t), \mathbf{u}(t), \boldsymbol{\theta}).
\end{equation}
Equation \eqref{eq:y_sep} expresses the algebraic states as a function of dynamic states, inputs, and system parameters, transforming the original DAE model into a more tractable dynamical system. This assumption is widely adopted in power system modeling \cite{tran2024solving}, and could be adopted into our DC modeling approach. Next, incorporating \eqref{eq:y_sep} into \eqref{eq:gen_dyn_2} reduces the DAE to the following parametric nonlinear dynamical system:
\begin{subequations}
\begin{align}
\dot{\mathbf{x}}(t) &= \mathcal{F}(\mathbf{x}(t), \mathbf{u}(t); \boldsymbol{\theta}) \\
\mathbf{z}(t) &= \mathcal{H}(\mathbf{x}(t), \mathbf{u}(t); \boldsymbol{\theta})
\end{align}
\end{subequations}
where 
$\mathcal{F}(\mathbf{x}, \mathbf{u}; \boldsymbol{\theta}) := \mathbf{f}(\mathbf{x}, \tilde{\mathbf{g}}(\mathbf{x}, \mathbf{u}, \boldsymbol{\theta}), \mathbf{u}, \boldsymbol{\theta})$ 
and $\mathcal{H}(\mathbf{x},\mathbf{u}(t); \boldsymbol{\theta})$
 $:= \mathbf{h}(\mathbf{x}, \tilde{\mathbf{g}}(\mathbf{x}, \mathbf{u}, \boldsymbol{\theta}), \boldsymbol{\theta})$.
 These reduced dynamics can be further simplified by moving the deterministic exogenous influences into the parameter vector and the system state. Specifically, the droop controller setpoints $(P_\text{PCC}^\text{ref}, Q_\text{PCC}^\text{ref}, V_\text{PCC}^\text{ref}, f_\text{PCC}^\text{ref})$ are fixed before the deployment time and remain constant throughout the simulation horizon. Therefore, it is justified to treat them as system parameters. The load profile $\{P_L(t), Q_L(t)\}$ 
evolves according to the parameterized profiles defined in Section \ref{sec:EMT_cases} (step and ramp), whose generating parameters (e.g., step magnitude, ramp duration, surge instant) are also encoded in $\boldsymbol{\theta}$. The breaker states 
$(s^\text{CB}_\text{grid}(t), s^\text{CB}_\text{inv}(t))$, which govern mode-switching transitions, are likewise generated by a deterministic switching law parameterized by the pre-scheduled switching instant $t_s$ and the pre-switching and post-switching mode 
configurations, both of which are captured in $\boldsymbol{\theta}$. These simplifications further reduce the DC EMT dynamics to the autonomous parametric dynamical system:
\begin{equation}
\label{eq:comp_dyn}
\dot{\mathbf{z}}(t) = \mathcal{G}(\mathbf{z}(t); \boldsymbol{\theta}),
\end{equation}
where $\mathcal{G}(\cdot)$ is the EMT compact autonomous dynamics that encapsulates the combined effect of the underlying DAE dynamics, the droop control law, and the deterministic load and circuit breaker profiles. From section \ref{sec:DAE}, we recall that
$\mathbf{z}(t) = [V_\text{abc}^\text{PCC}(t), I_\text{abc}^\text{PCC}(t)]$. This formulation reflects the limited observability encountered in practice. While circuit-level components are fully observable in white-box simulation models, typical IBR controllers remain black-box, and field measurements are generally restricted to a few monitored nodes \cite{isone_op22} such as the PCC. The proposed framework, which relies only on observable PCC waveforms, is therefore consistent with both settings.

As described in the Section \ref{sec:EMT_cases}, the IT load may experience a step-change, a gradual ramp, or a mode-switching during the DC operation. From a control theory perspective, these instances can be viewed as realizations of the autonomous dynamical system \eqref{eq:comp_dyn} under parametric uncertainty in $\boldsymbol{\theta}$. Constructing an adaptive digital twin, therefore, boils down to solving a system identification problem \eqref{eq:sysid_task} as described in the next section.

\section{ICL-based Adaptive Digital Twin}

This section presents an ICL-based framework that forms the core of the proposed adaptive digital twin. We first give a formal definition of an EMT-level digital twin, followed by formulating the problem as a system identification task under parametric uncertainty. Next, we provide background on ICL, and finally describe the training, finetuning, and evaluation procedures of the proposed digital twin as illustrated in Fig.~\ref{fig:AD_twin}.
\begin{definition}[EMT-level Digital Twin]
\label{def:digital_twin}
An EMT-level digital twin is a surrogate model 
$\widehat{\mathcal{M}}$ that approximates the discretized EMT dynamics $\mathcal{G}^{\Delta}(\cdot\,; \boldsymbol{\theta})$ of a physical DC facility under configuration $\boldsymbol{\theta}$. It maps an EMT state to its one-step-ahead state such that the predicted trajectory $\widehat{\mathbf{z}}(t)$ closely matches the recorded EMT response $\mathbf{z}(t)$ of the underlying high-fidelity simulator. The twin is termed \emph{adaptive} if it can adjust its predictions to an unseen configuration 
$\boldsymbol{\theta}_{\text{test}}$ at inference time, without retraining its parameters.
\end{definition}
\subsection{Inference-time System Identification}
In Section \ref{sec:prob_form}, we have established that EMT dynamics at the PCC are governed by the autonomous parametric dynamical system \eqref{eq:comp_dyn}. The nonlinear map $\mathcal{G}(\cdot\,; 
\boldsymbol{\theta})$  captures the collective effect of the underlying DAE dynamics, droop control laws, load profiles, and circuit breaker schedules. Accurate estimation of the map $\mathcal{G}(\cdot\,; 
\boldsymbol{\theta})$ is critical for planning, maintenance, and reliability assessment of the DC facility. However, obtaining a closed-form representation of $\mathcal{G}$ is extremely challenging, owing to the large-scale nature of modern DCs and their complex interactions with the upstream power grid. In practice, high-fidelity EMT simulation is adopted to perform what-if analysis and take corrective measures. Despite their accuracy, commercial EMT solvers like PSCAD and EMTP-RV are computationally slow, and require substantial time to generate EMT responses across a broad range of DC configurations, thereby limiting their practicality for large-scale planning studies. To circumvent this issue, a digital twin can be employed which achieves orders-of-magnitude faster inference than EMT solvers\cite{ogiesoba2026real}. However, there is a key limitation of such digital twins: \textit{inability to adapt to new configurational or operational DC instances}. Simply put, existing digital twins train on a preset power system configuration, and produce an incorrect EMT response once the configuration changes.  To address this challenge, we propose an adaptive digital twin, which is posed as the following inference-time system identification problem:
\begin{equation}
\text{Approximate} \;\; \mathcal{G}(\cdot\,; \boldsymbol{\theta}_{\text{test}}) 
\quad \text{for an unknown} \;\; \boldsymbol{\theta}_{\text{test}} 
\sim \mathcal{P}_\Theta,
\label{eq:sysid_task}
\end{equation}
where $\mathcal{P}_\Theta$ is a distribution over the configuration space $\Theta$ that characterizes the stochasticity of the configurations and operating regimes of interest. The unknown parameter $\boldsymbol{\theta}_{\text{test}}$ is not directly observable; instead, the surrogate must recover the underlying dynamics from a limited set of EMT samples generated under $\boldsymbol{\theta}_{\text{test}}$.
This requirement of parameter-agnostic, sample-efficient adaptation at inference time motivates the ICL framework that is introduced in the following subsection.
\subsection{Background on In-Context Learning}
\label{sec:icl_bg}
The ICL model implicitly identifies the underlying task structure from the in-context examples and utilizes this information to predict an output for the query input. Formally, let $\{\mathcal{T}_{\boldsymbol{\vartheta}}\}_{\boldsymbol{\vartheta} \in \Xi}$ be a family of tasks indexed by a parameter $\boldsymbol{\vartheta}$, where each task $\mathcal{T}_{\boldsymbol{\vartheta}}$ defines an input-output mapping $f_{\boldsymbol{\vartheta}}: \mathcal{X} \to \mathcal{Y}$. 
A parametrized ICL model $\mathcal{M}_{\phi}$ takes as input a sequence of in-context pairs $\{(x_j, f_{\boldsymbol{\vartheta}}(x_j))\}_{j=1}^{n}$ 
drawn from a task $\mathcal{T}_{\boldsymbol{\vartheta}}$, followed 
by a query input $x_{n+1}$, and produces a prediction:
\begin{equation}
\widehat{y}_{n+1} = \mathcal{M}_{\phi}\big(x_{n+1}; \, 
\{(x_j, f_{\boldsymbol{\vartheta}}(x_j))\}_{j=1}^{n}\big).
\label{eq:icl_def}
\end{equation}
The model  $\mathcal{M}_{\phi}$ is meta-trained over the task distribution 
$\mathcal{P}_\vartheta$ to minimize prediction error across both in-context samples and the query examples drawn from diverse tasks. As a result, it can generalize to unseen tasks $\mathcal{T}_{\boldsymbol{\vartheta}^*}$ 
at inference time without explicit access to 
$\boldsymbol{\vartheta}^*$.

Recent advances in ICL have shown that such frameworks can be extended to control and power system applications like adaptive system identification \cite{forgione2023system}, adaptive voltage control \cite{jena2025llm}, unified adaptive and stable system identification \cite{jena2026lilad}. Inspired by these works, we adopt an ICL-based approach and meta-learn $\mathcal{G}(.;\boldsymbol{\theta})$ over the space of DC configurations $\boldsymbol{\theta} \in \Theta$. In the proposed setting, the in-context examples 
correspond to a short observed post-disturbance EMT trajectory generated under the unknown test configuration $\boldsymbol{\theta}_{\text{test}}$, and the query corresponds to an initial condition for which the future EMT trajectory is predicted. In the remainder of this section, we describe the training, adaptation, and evaluation procedures, followed by the corresponding algorithmic pseudocodes.

\subsection{Training, Adaptation, and Evaluation of the ICL-Based Digital Twin}
We instantiate the ICL framework of Section \ref{sec:icl_bg} for the EMT system identification task \eqref{eq:sysid_task} using a decoder-only transformer with a GPT-2 backbone \cite{radford2019language}. The proposed model is trained offline on a multi-task data pool consisting of EMT trajectories spanning diverse DC configurations. Once the training is complete, the ICL model is deployed online to predict EMT responses under new and previously unseen configurations via in-context adaptation. 

\subsubsection{Adaptive Digital Twin Set-up}
We execute the PSCAD EMT model of the data center described in Fig. \ref{fig:DC_model} across $M$ distinct configurations, and construct the following multi-task EMT data pool:
\begin{equation}
\mathcal{D}^{\text{train}}
:=
\{\mathcal{D}_i\}_{i=1}^{M}
=
\Big\{\{(\mathbf{z}_{i,j}, \mathcal{G}^{\Delta}(\mathbf{z}_{i,j}; \boldsymbol{\theta}_i))\}_{j=1}^{N}\Big\}_{i=1}^{M},
\end{equation}
where $\mathbf{z}_{i,j}$ denotes a sampled EMT state vector; $\mathcal{G}^{\Delta}(\mathbf{z}_{i,j}; \theta_i)$ is the discretized counterpart of the continuous non-linear map $\mathcal{G}$ defined in \eqref{eq:comp_dyn}; $\mathcal{D}_i$ is the EMT dataset for the $i-$th DC configuration; $M$ is the total number of DC configurations (equivalently,  total number of tasks); and $N$ is the number of samples within a task. Similarly, the test-time dataset is defined as the following:
\begin{equation}
\mathcal{D}^{\text{test}}
:=
\mathcal{D}_{M+1}
=
\{(\mathbf{z}_{M+1,j}, \mathcal{G}^{\Delta}(\mathbf{z}_{M+1,j}; \boldsymbol{\theta}_{M+1}))\}_{j=1}^{\bar{N}},
\end{equation}
where the test-time EMT dataset is assumed to be drawn from the $(M+1)-$th task with configuration $\boldsymbol{\theta}_{M+1}$, that has not been observed during training, and $\bar{N}$ is the number of test-time samples. 
\begin{remark}
    The state and one-step-ahead prediction pairs $(\mathbf{z}_{i,j}, \mathcal{G}^{\Delta}(\mathbf{z}_{i,j}; \boldsymbol{\theta}_i))$ within each task are randomly shuffled to remove sequential dependence between consecutive pairs. Therefore, $\mathbf{z}_{i,j}$ and $\mathbf{z}_{i, j+1}$ are not temporally adjacent states but rather independently drawn samples from the task-specific data pool $\mathcal{D}_i$.  
\end{remark}

We now introduce the following context-query sequence construction to help establish the adaptive digital twin framework.
\begin{subequations}
\label{eq:prompt_const}
\begin{align}
\mathscr{P}^i 
&:= \big\{\mathbf{z}_{i,1}, \mathcal{G}^{\Delta}(\mathbf{z}_{i,1}; \boldsymbol{\theta}_i), \ldots, 
\mathbf{z}_{i,n}, \mathcal{G}^{\Delta}(\mathbf{z}_{i,n}; \boldsymbol{\theta}_i), \mathbf{z}_{i,n+1}\big\}, \\
\mathscr{P}^i_j 
&:= \big\{\mathbf{z}_{i,1}, \mathcal{G}^{\Delta}(\mathbf{z}_{i,1}; \boldsymbol{\theta}_i), \ldots, 
\mathbf{z}_{i,j}, \mathcal{G}^{\Delta}(\mathbf{z}_{i,j}; \boldsymbol{\theta}_i), \mathbf{z}_{i,j+1}\big\}, \\
\mathscr{C}^i 
&:= \mathscr{P}^i \setminus \{\mathbf{z}_{i,n+1}\}, 
\quad 
\mathscr{C}^i_j := \mathscr{P}^i_j \setminus \{\mathbf{z}_{i,j+1}\},
\end{align}
\end{subequations}
where $\mathscr{P}^i$ denotes the full EMT prompt for the $i-$th task; $\mathscr{P}^i_j$ denotes the $j-$th prompt prefix; and $\mathscr{C}^i, \mathscr{C}^i_j$ denote the corresponding context and context prefix obtained by removing the trailing query state. The ICL model $\mathcal{M}_\phi$ conditions on $\mathscr{C}^i$ to predict the next state, which predicts the discretized one-step map evaluated at the query state $\mathbf{z}_{i,n+1}$. Formally, 
\begin{equation}
\widehat{\mathcal{G}}^{\Delta}(\mathbf{z}_{i,n+1}; \boldsymbol{\theta}_i) 
:= \mathcal{M}_{\phi}\big(\mathbf{z}_{i,n+1} \mid \mathscr{C}^i\big),
\end{equation}
where $\widehat{\mathcal{G}}^{\Delta}(\mathbf{z}_{i,n+1}; \boldsymbol{\theta}_i)$ is the ICL-predicted next state given the current state $\mathbf{z}_{i,n+1}$. Note that the same relation holds for any prompt prefix $\mathscr{P}^i_j$.
\subsubsection{In-context Training}
The ICL model $\mathcal{M}_\phi$ is meta-trained on the multi-task data pool $\mathcal{D}^{\text{train}}$ to learn the parametric map $\mathcal{G}^\Delta(\cdot\,; \boldsymbol{\theta})$ as a function class indexed by the DC configuration $\boldsymbol{\theta} \in \Theta$. For each task $\mathcal{D}_i$, we construct a randomly shuffled prompt $\mathscr{P}^i$ as explained in \eqref{eq:prompt_const}, and provide that to the ICL model. The model is trained to predict the one-step-ahead vector $\mathcal{G}^\Delta(\mathbf{z}_{i,j}; \boldsymbol{\theta}_i)$ 
at every position $j \in \{0, \ldots, n\}$, conditioned on the context prefix $\mathscr{C}^i_j$. In other words, the following relation holds for every prompt-prefix, $\widehat{\mathcal{G}}^\Delta(\mathbf{z}_{i,j}; \boldsymbol{\theta}_i) 
= \mathcal{M}_\phi\big(\mathbf{z}_{i,j} \mid \mathscr{C}^i_j\big), 
\quad j \in \{0, \ldots, n\}.$ The ICL training loss is a meta-objective function that minimizes the average prediction loss across all tasks and all prompt prefixes within each task.
\begin{eqnarray}
\label{eq:meta_loss}
&\mathcal{L}_{\text{ICL}}(\phi) := 
\frac{1}{M(n+1)} \sum_{i=1}^{M} 
\sum_{j=0}^{n} 
\Big\| \mathcal{M}_\phi\big(\mathbf{z}_{i,j} \mid \mathscr{C}^i_j\big) - \nonumber \\
& \hspace{5cm} \mathcal{G}^\Delta(\mathbf{z}_{i,j}; \boldsymbol{\theta}_i) \Big\|_2^2,
\end{eqnarray}
where $\|.\|$ is the standard $\ell_2$ norm. Here, the inner sum is the per-task averaged prediction error over the $n+1$ context-query pairs, and the outer sum aggregates this error over $M$ training tasks. Each gradient step on \eqref{eq:meta_loss} thus jointly improves the model's predictive accuracy on the current query position and its ability to extract task-specific dynamics from the preceding context.

A key property of this training procedure is that the 
configuration $\boldsymbol{\theta}_i$ is never provided to the model as an explicit input; the model receives only the shuffled (state, next-state) pairs from task $\mathcal{D}_i$. Through meta-training over a diverse distribution of tasks, $\mathcal{M}_\phi$ learns to identify the underlying configuration implicitly from the context and predict the 
corresponding $\mathcal{G}^\Delta(\cdot\,; \boldsymbol{\theta})$ mapping accordingly. The trained model $\mathcal{M}_{\phi^*}$ therefore acts as an inference-time system identification algorithm in the sense of \cite{garg2022can,forgione2023system}, encoding the ability to recover the dynamics of an unseen task from a short context. The ICL training procedure is summarized in  Algorithm \ref{algor:ADT_train}.
\begin{algorithm}[t]
    \begin{algorithmic}[1]
\caption{ Adaptive Digital Twin In-context Training}
\label{algor:ADT_train}
    \STATE {\textbf{Input:}} Offline EMT prompt set $\{\mathscr{P}_i\}_{i=1}^{M}$, with each $\mathscr{P}_i$ comprising $N$ state and one-step-ahead prediction pairs; learning rate $\eta$;batch size $B$; and total iterations $\mathcal{K}$.
    \STATE {\textbf{Initialize:}} Adaptive digital twin parameters $\phi^{(0)}$
     \FOR{$k = 0, \hdots, \mathcal{K}-1$}
     \STATE Sample a random batch of $B$ prompt prefixes $\{\mathscr{P}_{B}^i\}_{i=1}^M$ where $B<N$.
     \STATE Compute the ICL training loss $\mathcal{L}_{\text{ICL}}(\phi)$ as described in \eqref{eq:meta_loss} over the mini batch $\{\mathscr{P}_{B}^i\}_{i=1}^M$.
     \STATE Update $\phi^{(k+1)} \leftarrow \phi^{(k)} - \eta \nabla_\phi \mathcal{L}_{\text{ICL}}(\phi^{(k)})$.
     \ENDFOR
  \STATE {\textbf{Output:}} Optimal parameters $\phi^{*} = \phi^{(\mathcal{K})}$
\end{algorithmic}
\end{algorithm}
\begin{algorithm}
\caption{Adaptive Digital Twin In-Context Adaptation and Evaluation}
\label{algor:ADT_eval}
\begin{algorithmic}[1]
\STATE \textbf{Input:} Pre-trained parameters $\phi^{*}$; test-time 
EMT dataset $\mathcal{D}^{\text{test}}$ collected under unseen 
configuration $\boldsymbol{\theta}_{M+1}$, comprising $\bar{N}$ 
temporally ordered samples; initial condition $\mathbf{z}_0$; 
prediction horizon $K$.
\STATE \textbf{Initialize:} Construct the test-time context 
$\mathscr{C}^{M+1}$ from $\mathcal{D}^{\text{test}}$ in its 
original temporal order.
\STATE Set $\widehat{\mathbf{z}}_{M+1, 0} \leftarrow \mathbf{z}_0$.
\FOR{$k = 1, \ldots, K$}
\STATE Predict next state: 
$\widehat{\mathbf{z}}_{M+1, k} \leftarrow \mathcal{M}_{\phi^*}\big(\widehat{\mathbf{z}}_{M+1, k-1} \mid \mathscr{C}^{M+1}\big)$.
\ENDFOR
\STATE \textbf{Output:} Predicted EMT trajectory 
$\{\widehat{\mathbf{z}}_{M+1, k}\}_{k=1}^{K}$.
\end{algorithmic}
\end{algorithm}

\subsubsection{In-Context Adaptation and Evaluation}
During the deployment phase, the trained model $\mathcal{M}_{\phi^*}$ is supplied with a test-time dataset $\mathcal{D}^{\text{test}}$ collected from a data center operating under the unseen 
configuration $\boldsymbol{\theta}_{M+1}$ consisting of $\bar{N}$ EMT samples. Specifically, we execute an EMT simulation initialized at a post-disturbance time instance and record a short prefix of the resulting transient to create $\mathcal{D}^{\text{test}}$.  In contrast to the offline training 
phase, $\mathcal{D}^{\text{test}}$ is \emph{not} shuffled; instead, the EMT sequence is fed to the ICL model in its original temporal order. This choice provides $\mathcal{M}_{\phi^*}$ with 
additional temporal information about the post-disturbance dynamics, which improves predictive accuracy. We note, however, 
that the model still produces near-accurate EMT predictions when the test-time pairs are shuffled, indicating that the benefit of temporal ordering is incremental rather than essential. Formally,  $\mathcal{D}^{\text{test}}$ is transformed into the following  test-time context:
\begin{eqnarray}
    &\mathscr{C}^{M+1} := \big\{\mathbf{z}_{M+1,1}, \mathcal{G}^{\Delta}(\mathbf{z}_{M+1,1}; \boldsymbol{\theta}_{M+1}) \nonumber\\
     &\hspace{2cm}\mathbf{z}_{M+1,n}, \mathcal{G}^{\Delta}(\mathbf{z}_{M+1,n}; \boldsymbol{\theta}_{M+1})\big\}, 
\end{eqnarray}
where consecutive samples $\mathbf{z}_{M+1, j}$ and 
$\mathbf{z}_{M+1, j+1}$ are temporally adjacent EMT state vectors from the underlying simulation. 

The adaptation and evaluation phases are then performed jointly through recursive prediction. Given the test-time context $\mathscr{C}^{M+1}$  and an initial condition $\mathbf{z}_0$, the 
model $\mathcal{M}_{\phi^*}$ generates the EMT trajectory auto-regressively by treating its previous prediction as the query for the next step:
\begin{equation}
\widehat{\mathbf{z}}_{M+1, k} = 
\mathcal{M}_{\phi^*}\big(\widehat{\mathbf{z}}_{M+1, k-1} \mid \mathscr{C}^{M+1}\big), 
\quad \widehat{\mathbf{z}}_{M+1, 0} = \mathbf{z}_0,
\end{equation}
where, $\widehat{\mathbf{z}}_{M+1, k}$ denotes the predicted EMT state for time instance $k$ and $\mathbf{z}_{0}$ is the initial condition for EMT prediction. The recursive rollout 
continues until the desired prediction horizon is reached. Predictive accuracy is then assessed via mean absolute error (MAE) and root-mean-squared error (RMSE) computed against the ground-truth PSCAD-simulated trajectory. We present the adaptation and evaluation algorithm in Algorithm \ref{algor:ADT_eval}
\subsubsection{Practical Feasibility of Test-Time Data Acquisition}
A critical concern with the proposed framework is the operational viability of acquiring the test-time data $\mathcal{D}^{\text{test}}$ from a real-world DC facility. We propose a deployment pathway that addresses this concern in the following Remark \ref{remark:feasibility}.

\begin{remark}
\label{remark:feasibility}
During routine operation, point-on-wave (POW) recording devices installed at the PCC continuously monitor the transient response of the variables of interest following any disturbance, such as a sudden change in IT load. For any new configurational or operational instance, these recordings can be incrementally appended to $\mathcal{D}^{\text{test}}$, which keeps gathering sequential EMT samples until a minimum size $\bar{N}$ is reached. The digital twin stays idle during the accumulation phase, introducing a deployment \emph{latency} before prediction can 
begin. Once the minimum sample size is attained, $\mathcal{D}^{\text{test}}$ is assembled into the test-time context $\mathscr{C}^{M+1}$, and the trained digital twin conditions on this context to perform recursive one-step-ahead prediction over the desired horizon. In practice, the test-time sample size is significantly smaller than the training-time per-task sample size ($\bar{N} \ll N$), which makes the test-time data acquisition both feasible and efficient. 
\end{remark}
\section{Case Studies}
In this section, we evaluate the proposed ICL-based twin on three EMT scenarios: (i) step-change-in-load-induced EMT, (ii) gradual-change-in-load-induced EMT, (iii) mode-switching-induced EMT, as described in Section \ref{sec:EMT_cases}. We first introduce the baselines, then describe the simulation setup, and finally present a detailed performance comparison.
\subsection{Baselines}
 We employ three non-adaptive baselines to compare the performance of the proposed framework. All baselines are task-specific and are trained on a larger dataset $\mathcal{D}^\text{test-full}$ collected from a single DC EMT scenario. Instead of adapting to new scenarios, each baseline predicts the EMT response from a new post-disturbance initial condition within the scenario on which it was trained. The baselines are described below.
\begin{itemize}
    \item \textbf{ARMAX}: The autoregressive moving-average model with exogenous inputs (ARMAX) is a classical system identification method \cite{liu2015armax}. As a fixed linear model, it provides an interpretable baseline.
    \item \textbf{LSTM}: The long short-term memory (LSTM) network is another widely-used surrogate modeling method in power systems \cite{cao2020scalable}, which offers a richer representational capacity over classical methods.
    \item \textbf{PINN}: The physics-informed neural network (PINN) augments the data-driven training objective with the governing physical equations of the system as a soft constraint.  Recent works \cite{huang2022PINN} have reported the effectiveness of PINN in power systems, owing to its ability to respect the physical nature
    of a power grid.
\end{itemize}
\subsection{Simulation Setup}
\label{sec:sim_setup}
We construct a high-fidelity EMT model of the DC power system as illustrated in Fig.~\ref{fig:DC_model} in PSCAD software. The upstream transmission grid, originally defined in PSSE format, is converted into a PSCAD-compatible EMT representation using the E-TRAN interface \cite{etran}. This interface extends a static positive-sequence network into a dynamic network and preserves the dynamics of other dynamic components in the EMT domain. This model serves as the ground-truth simulator for generating the training data pool $\mathcal{D}^{\text{train}}$, test-time data $\mathcal{D}^{\text{test}}$ for the adaptive digital twin, and a longer version of the test-time data $\mathcal{D}^{\text{test-full}}$ for the baselines. The model is simulated at an EMT timestep of $\Delta t = 5~\mu$s, producing the sub-cycle voltage and current dynamics at the PCC. As described in the mathematical model \eqref{eq:comp_dyn}, the three-phase voltage and current waveforms 
$\mathbf{z}(t) = [V_\text{abc}^\text{PCC}(t), I_\text{abc}^\text{PCC}(t)]$ are recorded as the EMT outputs of interest.
\subsubsection{PSCAD Model Configuration}
The upstream transmission grid is modeled as the IEEE 39-bus benchmark system \cite{pscad_ieee39} interfaced to the DC facility through a 100-MVA interface transformer with 0.1~p.u. leakage reactance and 0.01~p.u. winding resistance, configured in a grounded-wye/grounded-wye arrangement. During islanded operation, the DC receives power from a grid-forming inverter backed by a battery storage unit, where the battery is modeled as an ideal \textit{DC} source under the infinite-capacity assumption as stated in Assumption \ref{assum:infinite_dc}. The aggregate IT load is modeled as a constant-power (PQ) load at the PCC, with a time-varying power setpoint. This load is connected to the PCC, and is driven by the disturbance scenarios presented in Section \ref{sec:EMT_cases}. The nominal model parameters shared across all cases are summarized in Table~\ref{tab:pscad_params}, and the configuration parameters specific to each case are listed in Table~\ref{tab:case_configs}.

\subsubsection{Data Generation Pipeline}
We generate the multi-configuration EMT data pool by simulating the PSCAD model across $M$ distinct configurations. Each configuration is associated with a unique parameter vector $\boldsymbol{\theta}_i$, obtained by varying key DC parameters, such as load power factors and apparent power rating, over the ranges listed in Table ~\ref{tab:case_configs}. For each configuration, the EMT response to the disturbance (e.g., a step change or gradual ramp in load) is simulated, and the resulting three-phase voltage and current waveforms $\mathbf{z}(t) = [V_\text{abc}^\text{PCC}(t), I_\text{abc}^\text{PCC}(t)]$ are recorded at the PCC over the simulation horizon. Each configuration thus yields a 
task-specific dataset $\mathcal{D}_i$ of $N$ state and one-step-ahead prediction pairs. For evaluation, we hold out an unseen test configuration $\boldsymbol{\theta}_{M+1}$ and generate two datasets under it: a small test-time dataset $\mathcal{D}^{\text{test}}$ of 
$\bar{N}$ samples for the proposed adaptive digital twin, and a larger dataset 
$\mathcal{D}^{\text{test-full}}$ for task-specific training of the non-adaptive baselines. We report the values of $M$, $N$ and $\bar{N}$ in Table \ref{tab:icl_params}.

\begin{table}[t]
\centering
\caption{Nominal PSCAD Model Parameters (Common to All Cases)}
\label{tab:pscad_params}
\begin{tabular}{lcc}
\toprule
\textbf{Parameter} & \textbf{Symbol} & \textbf{Value} \\
\midrule
EMT simulation timestep    & $\Delta t_{\text{sim}}$ & 5 $\mu$s \\
Nominal grid voltage       & $V_\text{grid}$       & 345 kV \\
Transformer rating         & $S_T$            & 100 MVA \\
Nominal PCC frequency      & $f_\text{PCC}^\text{ref}$  & 60 Hz \\
Nominal PCC voltage        & $V_\text{PCC}^\text{ref}$  & 1.0 p.u. \\
$P$--$f$ droop gain        & $K_p$            & 0.0005 \\
$Q$--$V$ droop gain        & $K_q$            & 0.0005 \\
Filter time constant       & $T$              & 0.2 s \\
Measurement gain           & $G$              & 1.0 \\
Actuator output impedance  & $Z_\text{act}$        & 1.0 $\Omega$ \\
Base power                 & $S_\text{base}$       & 100 MVA \\
Base voltage (L--L)        & $V_\text{base}$       & 345 kV \\
Base current               & $I_\text{base}$       & 167.3 A \\
\bottomrule
\end{tabular}
\end{table}
\begin{table}[t]
\centering
\caption{Configuration Parameters for the Three EMT Case Studies}
\label{tab:case_configs}
\begin{threeparttable}
\small
\setlength{\tabcolsep}{3pt}
\begin{tabular}{@{}lllrr@{}}
\toprule
\textbf{Case} & 
\makecell[l]{\textbf{Dist.}\\\textbf{Type}} & 
\makecell[l]{\textbf{Stoch.}\\\textbf{Param.}} & 
\makecell[r]{\textbf{Nominal}\\\textbf{Value}} & 
\makecell[r]{\textbf{Variation}\\\textbf{Range}} \\
\midrule
\multirow{2}{*}{\makecell[l]{Step-\\change}} & 
\multirow{2}{*}{\makecell[l]{Step\\load}} & 
$P_L$ & 25\,MW & {[}0, 320{]}\,MW \\
& & $Q_L$ & 8.44\,MVAR & {[}0, 105.2{]}\,MVAR \\
\midrule
\multirow{2}{*}{\makecell[l]{Gradual-\\change}} & 
\multirow{2}{*}{\makecell[l]{Ramp\\load}} & 
$P_L$ & 25\,MW & {[}0, 320{]}\,MW \\
& & $Q_L$ & 8.44\,MVAR & {[}0, 105.2{]}\,MVAR \\
\midrule
\makecell[l]{Mode-\\switch} & 
\makecell[l]{Breaker\\toggle} & 
$S_{GFM}$ & 150\,MVA & {[}$-$80, 80{]}\,MVA \\
\bottomrule
\end{tabular}
\begin{tablenotes}
\footnotesize
\item[$\dagger$] Event timing: $t_s = 0.5$\,s for all cases; 
ramp duration $\Delta t = 0.05$\,s for gradual-change.
\item[$\ddagger$] $S_{GFM}$ is the GFM inverter apparent power; 
its range reflects the dispatch considered across configurations.
\end{tablenotes}
\end{threeparttable}
\end{table}

\begin{table}[t]
\centering
\caption{ICL Model Architecture and Training Hyperparameters}
\label{tab:icl_params}
\begin{tabular}{lcc}
\toprule
\textbf{Parameter} & \textbf{Symbol} & \textbf{Value} \\
\midrule
\multicolumn{3}{l}{\textit{Model Architecture (GPT-2 backbone)}} \\
\midrule
Number of transformer layers   & $L$       & 6 \\
Number of attention heads       & $H$       & 4 \\
Embedding dimension             & $d$       & 128 \\
Context length                  & $n$       & 36 \\
\midrule
\multicolumn{3}{l}{\textit{Training and Test Configuration}} \\
\midrule
Number of training tasks        & $M$       & 1000 \\
Samples per task                & $N$       &  1600\\
Optimizer                       & --        & AdamW \\
Learning rate                   & $\eta$    & 5e-7 \\
Batch size                      & $B$       & 128 \\
Training iterations             & $\mathcal{K}$ & 300000 \\
Test-time context size          & $\bar{N}$ & 36 \\
\bottomrule
\end{tabular}
\end{table}
\begin{table*}[t]
\centering
\caption{Comparison of Three-Phase PCC Voltage Prediction 
Accuracy, Execution Speed, and Adaptivity Across Methods 
and Test Cases}
\label{tab:method_comparison}
\begin{threeparttable}
\begin{tabular}{lc cc cc cc cc}
\toprule
\multirow{2}{*}{\textbf{Method}} & \multirow{2}{*}{\textbf{Adapts?}} 
& \multicolumn{2}{c}{\textbf{Step-change}} 
& \multicolumn{2}{c}{\textbf{Gradual-change}} 
& \multicolumn{2}{c}{\textbf{Mode-switching}} 
& \multicolumn{2}{c}{\textbf{Avg. over all cases}} \\
\cmidrule(lr){3-4} \cmidrule(lr){5-6} \cmidrule(lr){7-8}
\cmidrule(lr){9-10}
 & & \textbf{MAE} & \textbf{RMSE} 
 & \textbf{MAE} & \textbf{RMSE} 
 & \textbf{MAE} & \textbf{RMSE} 
 & \textbf{ET (s)} & \textbf{Gain ($\times$)} \\
\midrule
PSCAD (ground truth) & -- & -- & -- & -- & -- & -- & -- 
& 128.06 & $1\times$ \\
\textbf{ICL (ours)} & $\checkmark$ 
& 0.00078 & 0.00316 
& 0.00152 & 0.00438 
& 0.00119 & 0.00524 
& 9.911 & 12.92$\times$ \\
ARMAX & $\times$ 
& 0.06613 & 0.09493 
&   0.05090 & 0.07071 
& 0.08273 & 0.10764 
& 0.026 & 4925$\times$ \\
LSTM & $\times$ 
& 0.00842 & 0.01343 
& 0.00067 & 0.00092 
& 0.00248 & 0.01201 
& 1.156 & 110.78$\times$ \\
PINN & $\times$ 
& 0.04139 & 0.04818 
& 0.03338 & 0.04204
& 0.01977 & 0.03332 
& 0.791 & 161.90$\times$ \\
\bottomrule
\end{tabular}
\begin{tablenotes}
\footnotesize
\item[$\dagger$] MAE and RMSE are computed over the 
three-phase PCC voltage $V_\text{abc}^\text{PCC}$.
\item[$\ddagger$] ET denotes execution time; Gain denotes 
speedup relative to full PSCAD simulation; both are 
averaged over the three EMT scenarios.
\end{tablenotes}
\end{threeparttable}
\end{table*}

\subsection{Case Study Results}
\label{sec:case_results}
We now present the prediction results of the proposed adaptive digital twin across the three representative EMT scenarios, compared with the baselines. In all cases, the test configuration $\boldsymbol{\theta}_{M+1}$ is unseen during training, and the adaptive digital twin relies 
solely on the small test-time context $\mathcal{D}^{\text{test}}$ to adapt, whereas the baselines are trained on the larger configuration-specific dataset $\mathcal{D}^{\text{test-full}}$.
\subsubsection{Case 1: Step-Change-in-Load-Induced EMT}
\begin{figure*}[t]
\centering
\begin{subfigure}[b]{0.31\textwidth}
    \centering
    \includegraphics[width=\textwidth]{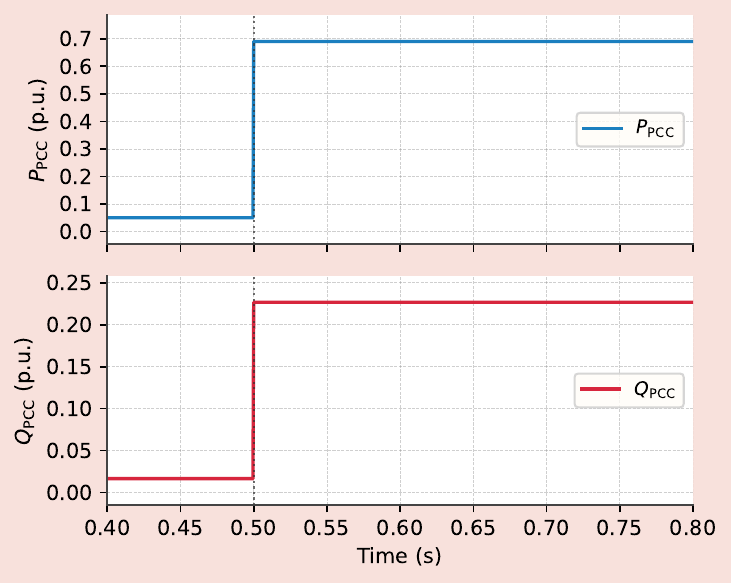}
    \caption{Step-change load profile ($P_L$, $Q_L$)}
    \label{fig:case1_pq}
\end{subfigure}
\hfill
\begin{subfigure}[b]{0.31\textwidth}
    \centering
    \includegraphics[width=\textwidth]{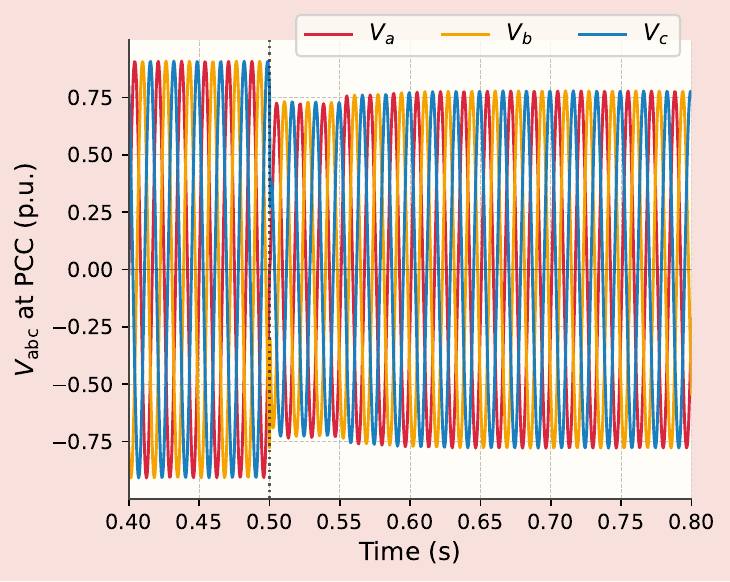}
    \caption{Ground-truth $V_\text{abc}^\text{PCC}$ (PSCAD)}
    \label{fig:case1_v_gt}
\end{subfigure}
\hfill
\begin{subfigure}[b]{0.31\textwidth}
    \centering
    \includegraphics[width=\textwidth]{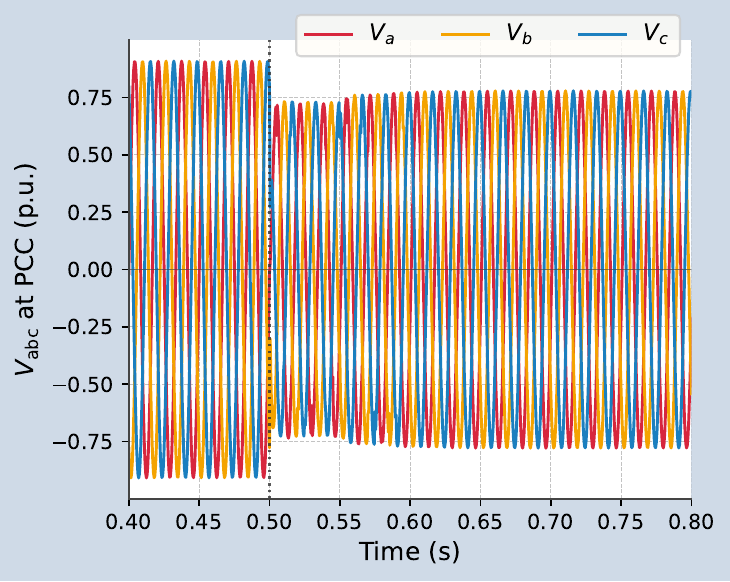}
    \caption{Predicted $V_\text{abc}^\text{PCC}$ (ICL-based adaptive DT)}
    \label{fig:case1_v_pred}
\end{subfigure}

\vspace{0.4em}

\begin{subfigure}[b]{0.31\textwidth}
    \centering
    \includegraphics[width=\textwidth]{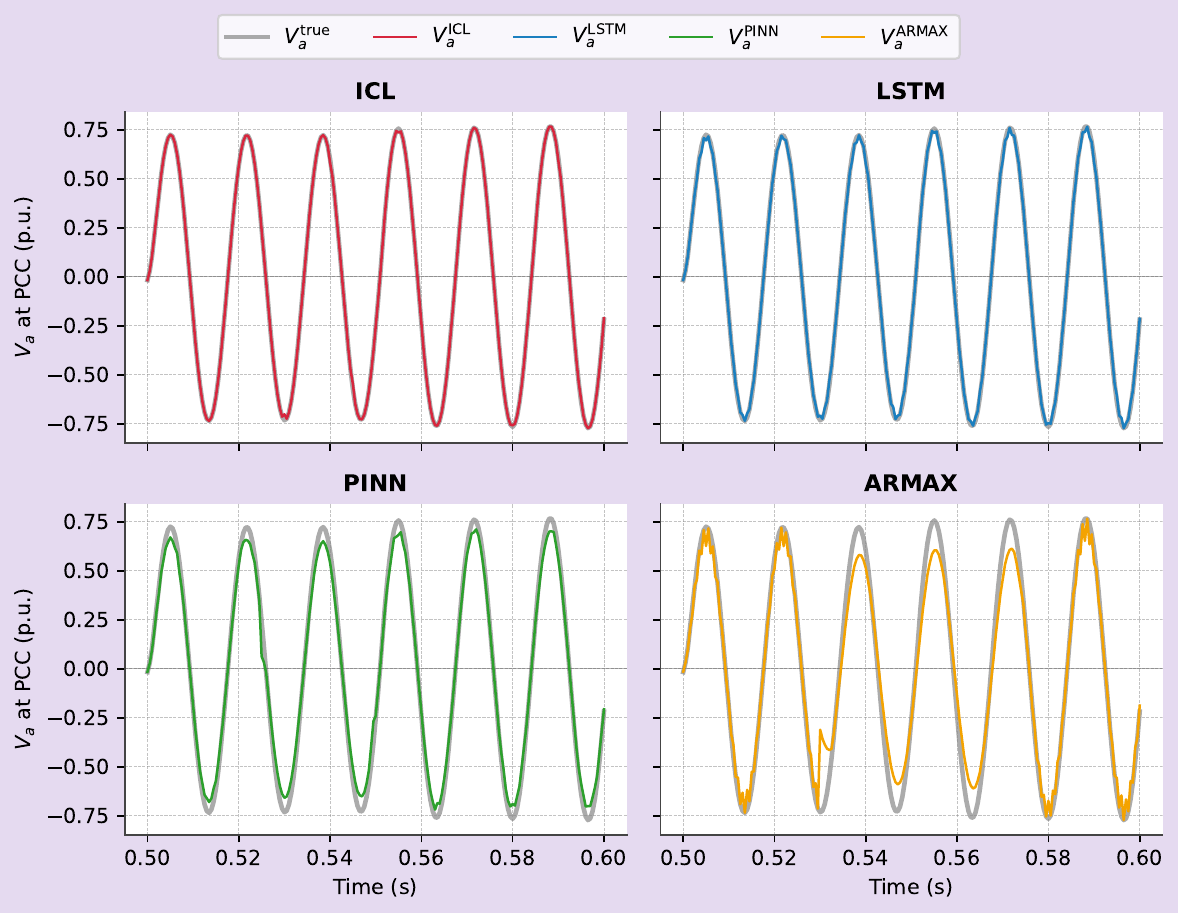}
    \caption{Case~1: $V_\text{a}^\text{PCC}$, all methods}
    \label{fig:case1_v_compare}
\end{subfigure}
\hfill
\begin{subfigure}[b]{0.31\textwidth}
    \centering
    \includegraphics[width=\textwidth]{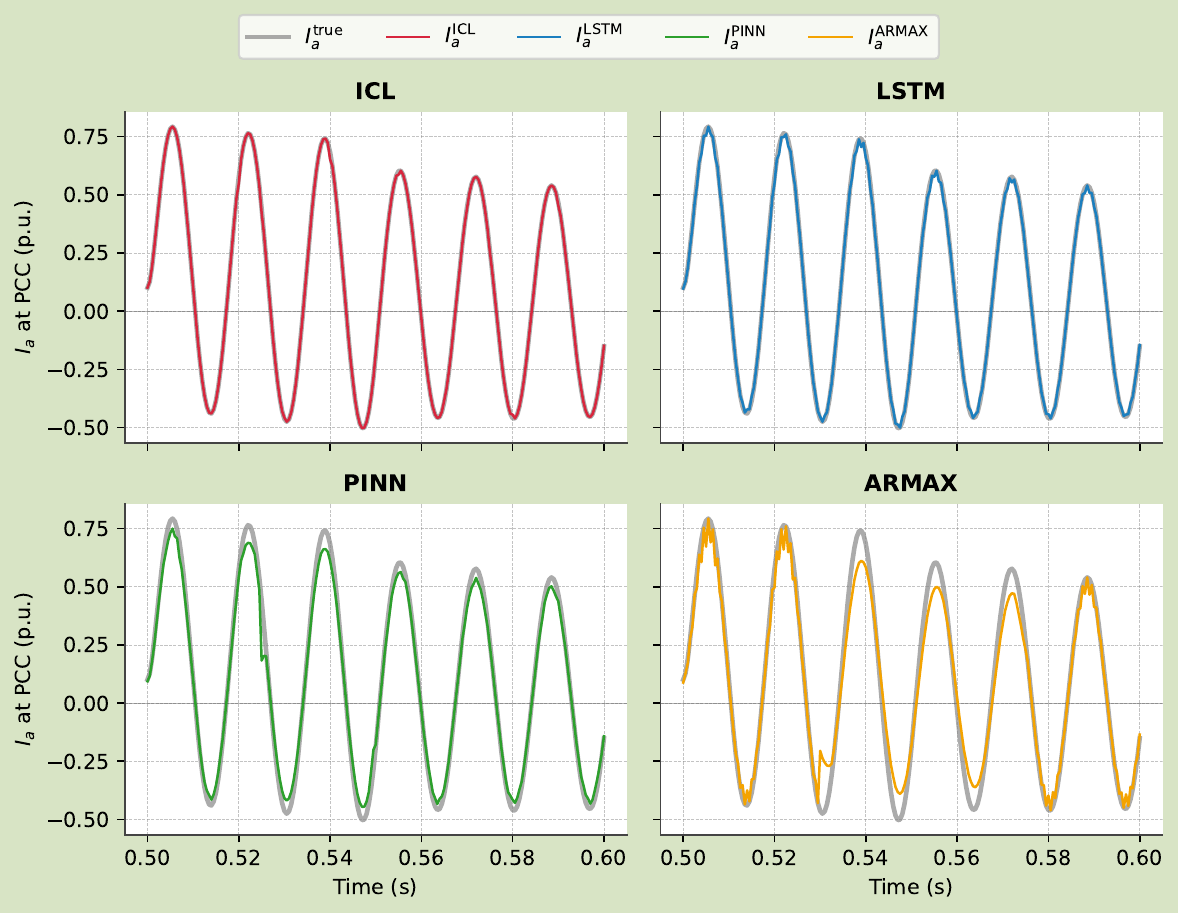}
    \caption{Case~1: $I_\text{a}^\text{PCC}$, all methods}
    \label{fig:case1_i_compare}
\end{subfigure}
\hfill
\begin{subfigure}[b]{0.31\textwidth}
    \centering
    \includegraphics[width=\textwidth]{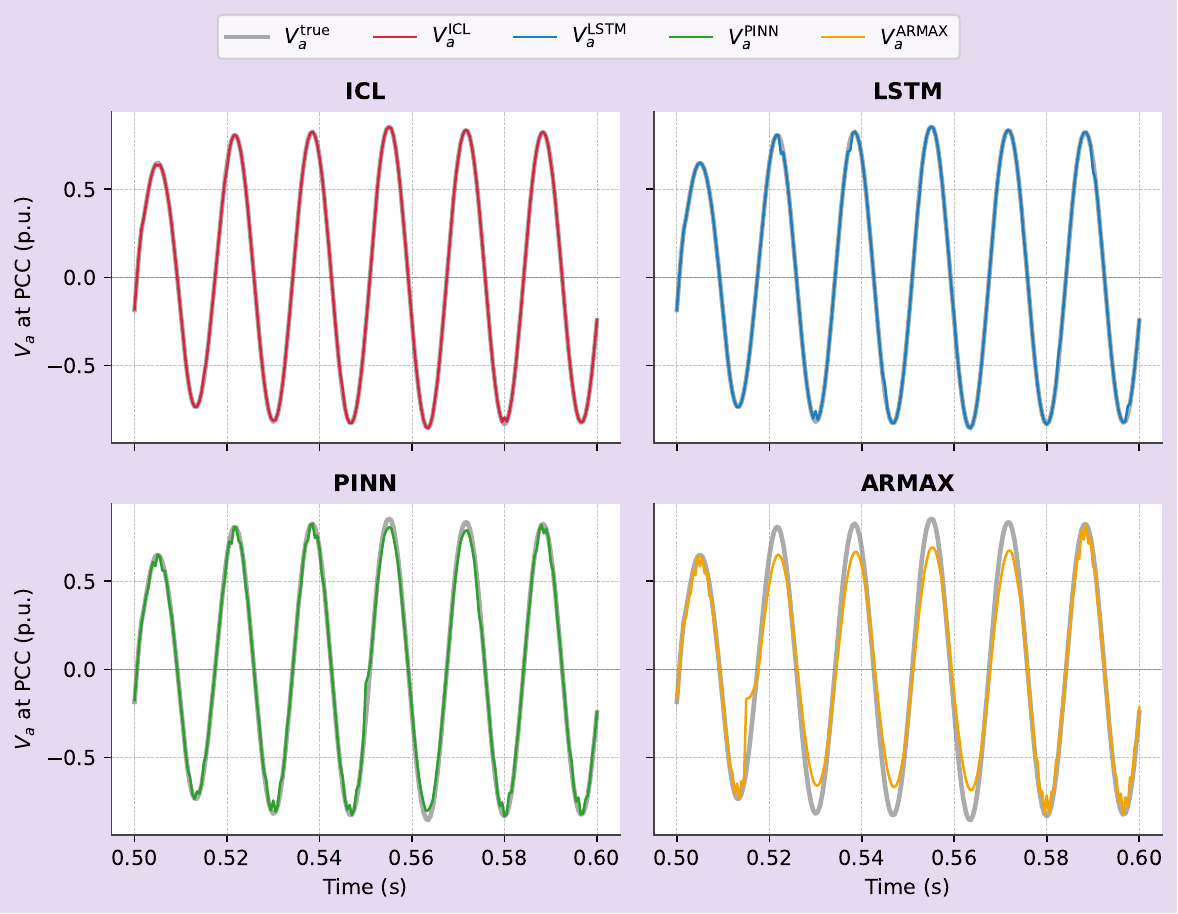}
    \caption{Case~3: $V_\text{a}^\text{PCC}$, all methods}
    \label{fig:case3_v_compare}
\end{subfigure}

\caption{\textit{EMT prediction results}. The top row presents the detailed Case~1 (step-change): (a) the applied load 
disturbance, (b) the ground-truth three-phase PCC voltage, and (c) the corresponding adaptive digital twin prediction. The 
bottom row presents single-phase cross-method comparisons: (d) PCC voltage and (e) PCC current for Case~1, and (f) PCC 
voltage for Case~3 (mode-switching). The proposed ICL-based adaptive digital twin accurately reproduces both the voltage and current 
EMT responses. The quantitative results for all cases, including Case~2 (gradual-change), are reported in 
Table~\ref{tab:method_comparison}.}
\label{fig:case_results}
\end{figure*}

In this scenario, the IT load undergoes an abrupt step change at $t_s = 0.5$\,s, as specified in Table~\ref{tab:case_configs}. Fig.~\ref{fig:case_results}(a) shows the applied disturbance in the active and reactive power profiles, while Figs.~\ref{fig:case_results}(b) and \ref{fig:case_results}(c) present the ground-truth and three-phase PCC voltage predicted by the adaptive digital twin, 
respectively. The proposed adaptive digital twin effectively reproduces the full three-phase voltage waveform, capturing both the post-disturbance transient and the subsequent steady-state sinusoidal waveform. Fig.~\ref{fig:case_results}(d) provides a single-phase cross-method comparison over a shorter transient window. The ICL-based twin closely tracks the ground-truth trajectory, achieving an MAE of $0.00078$ and RMSE of $0.00316$ 
(Table~\ref{tab:method_comparison}), which is roughly an order of magnitude lower than the LSTM and nearly two orders of magnitude lower than ARMAX and PINN. The classical ARMAX model, constrained 
by its linear structure, fails to capture the nonlinear transient behavior, while the PINN exhibits a visible mismatch during the fast post-disturbance interval. 

The current prediction follows the same trend as the voltage results. Fig.~\ref{fig:case_results}(e) presents the single-phase PCC current comparison across all methods, where the proposed 
twin again reproduces the ground-truth waveform with high fidelity, confirming that the framework accurately predicts both EMT outputs $\mathbf{z} = [V_\text{abc}^\text{PCC}, I_\text{abc}^\text{PCC}]$.

\subsubsection{Case 2: Gradual-Change-in-Load-Induced EMT}
Here, the IT load varies continuously over a ramp interval of $\Delta t = 0.05$\,s following the quadratic profile of \eqref{eq:gradual_load}, initiated at $t_s = 0.5$\,s. The smoother disturbance produces a milder transient than the 
step-change case. As reported in Table~\ref{tab:method_comparison}, the proposed adaptive digital twin attains an MAE of $0.00152$ 
and RMSE of $0.00438$, remaining highly accurate across the ramp. While the LSTM achieves a lower error on this 
particular configuration ($\text{MAE} = 0.00067$), it is restricted to the single configuration on which it was trained and lacks the inference-time adaptivity of the proposed twin. Both ARMAX and PINN exhibit substantially higher errors, consistent with the step-change case.

\subsubsection{Case 3: Mode-Switching-Induced EMT}
In this scenario, the DC transitions between 
grid-connected and islanded operation at $t_s = 0.5$\,s via a circuit breaker toggle, with the IT load held constant. This is the most challenging case, as the switching event induces a discontinuous change in the active current paths at the PCC. Fig.~\ref{fig:case_results}(f) presents the single-phase PCC voltage comparison across all methods. The proposed adaptive 
digital twin accurately captures the sharp mode-switching transient, achieving an MAE of $0.00119$ and RMSE of $0.00524$ (Table~\ref{tab:method_comparison}), markedly outperforming all three baselines. The non-adaptive baselines struggle with the abrupt topology change: ARMAX yields the highest error ($\text{MAE} = 0.08273$), while LSTM and PINN, though better, 
remain well above the proposed method. These results highlight the advantage of inference-time adaptation when the disturbance induces a structural change in the system dynamics.
\subsection{Computational Efficiency}
Table~\ref{tab:method_comparison} reports the execution time (ET) and the corresponding speedup (Gain) relative to full PSCAD simulation, averaged over the three scenarios. The proposed ICL-based digital twin achieves a $12.92\times$ speedup over the high-fidelity solver, reducing the per-scenario runtime from $128.06$\,s to $9.911$\,s. While the non-adaptive baselines are computationally faster, with ARMAX, LSTM, and PINN attaining speedups of $4925\times$, $110.78\times$, and $161.90\times$, respectively, this speed advantage stems from their simpler, single-configuration structure and comes at the cost of adaptivity. Each baseline must be retrained from scratch for every new DC configuration, whereas the proposed twin adapts to unseen configurations at inference time without 
retraining. Consequently, the modest runtime overhead of the ICL-based twin is a favorable trade-off. It delivers a substantial speedup over full EMT simulation while uniquely providing configuration-agnostic adaptation that the baselines fundamentally lack.
\section{Conclusion}
\label{sec:conclusion}
This paper presented an adaptive digital twin framework for EMT prediction in data centers based on in-context learning with large language models. By formulating EMT prediction as a system identification problem under parametric uncertainty, the proposed GPT-2-based framework adapts to unseen data center configurations at inference time using only a small set of EMT samples, without any retraining. Across three representative EMT scenarios, the adaptive twin achieved superior or comparable prediction accuracy relative to non-adaptive baselines while requiring substantially fewer test-time samples. In addition, it provided significant computational speedups over full EMT simulations in PSCAD. These results demonstrate the potential of ICL-based adaptive digital twins as scalable surrogates for EMT-level data center analysis.

Future work will focus on incorporating battery recharge dynamics, modeling converter-level behavior and harmonics within IT loads, reducing the adaptation latency associated with collecting post-disturbance samples, and extending the framework to grid-level impact studies, including voltage stability and converter interaction analysis.

\section*{Disclaimer}
The views expressed in this paper are the opinion of the authors and do not reflect the views of PJM Interconnection, L.L.C. or its Board of Managers of which Le Xie is a member.



\bibliographystyle{IEEEtran}
\bibliography{ref}

@electronic{pscad_ieee39,
  author       = {{Manitoba HVDC Research Centre}},
  title        = {IEEE 39 Bus System},
  journal      = {PSCAD Knowledge Base},
  url          = {https://pscad.com},
  year         = {2018},
  note         = {[Accessed: May 21, 2026]}
}

@article{tran2024solving,
  title={Solving differential-algebraic equations in power system dynamic analysis with quantum computing},
  author={Tran, Huynh TT and Nguyen, Hieu T and Vu, Long T and Ojetola, Samuel T},
  journal={Energy Conversion and Economics},
  volume={5},
  number={1},
  pages={28--41},
  year={2024},
  publisher={Wiley Online Library}
}

@article{ogiesoba2026real,
  title={Real-Time Surrogate Modeling for Fast Transient Prediction in Inverter-Based Microgrids Using CNN and LightGBM},
  author={Ogiesoba-Eguakun, Osasumwen Cedric and Ashenayi, Kaveh and Rath, Suman},
  journal={arXiv preprint arXiv:2603.29255},
  year={2026}
}

@techreport{iea2025energyai,
  author      = {{International Energy Agency}},
  title       = {Energy and {AI}},
  institution = {IEA},
  year        = {2025},
  address     = {Paris},
  url         = {www.iea.org/reports/energy-and-ai.},
  note        = {Accessed: 2026-05-27}
}

@misc{iea2025surge,
  author = {{International Energy Agency }},
  title  = {Data Centre Electricity Use Surged in 2025, Even with Tightening Bottlenecks Driving a Scramble for Solutions},
  year   = {2025},
  url = {www.iea.org/news.},
  note   = {[Online]. Accessed: 2026-05-27}
}

@techreport{isone_op22,
  author      = {{ISO New England}},
  title       = {{ISO New England Operating Procedure No.~22: Transmission Operations}},
  institution = {ISO New England Inc.},
  year        = {2024},
  note        = {Accessed: 2026-06-16}
}

@article{guest2023speeding,
  author  = {Guest, James and Commerford, Ian and Modi, Nilesh and Saadati, Sheler and Alonso, Juan Carlos and Kahingala, Thisandu},
  title   = {Speeding Up Electromagnetic Transient Simulations for Inverter-Based Resources: {Australian} Experience},
  journal = {IEEE Electrification Magazine},
  volume  = {11},
  number  = {3},
  pages   = {72--80},
  year    = {2023},
  month   = sep,
  doi     = {10.1109/MELE.2023.3291270}
}

@article{li2026ups,
  title={Coordinating multiple UPS batteries in datacenter for load flexibility and cost reduction},
  author={Li, Lunlong and Pu, Yang and Ju, Yi and Hu, Ziqi and Wang, Zhe},
  journal={Energy and Buildings},
  pages={117282},
  year={2026},
  publisher={Elsevier}
}

@inproceedings{ai2023vfd,
  title={Energy-Saving Practice of Variable Frequency Drive Technology in Data Center HVAC System},
  author={Ai, Chuanxian and Yang, Jinpeng and Chen, Rui},
  booktitle={2023 5th International Conference on Power and Energy Technology (ICPET)},
  pages={240--247},
  year={2023},
  organization={IEEE}
}

@inproceedings{seshmasetti2025review,
  title={A Review of Power System Challenges Stemming from Large AI Data Center Loads},
  author={Seshmasetti, Venkateswara R and Ismail, Abdulla and Khalid, Haris M},
  booktitle={2025 IEEE PES Conference on Innovative Smart Grid Technologies-Middle East (ISGT Middle East)},
  pages={1--5},
  year={2025},
  organization={IEEE}
}

@article{aidc2025grid,
  title={Electricity demand and grid impacts of AI data centers: Challenges and prospects},
  author={Chen, Xin and Wang, Xiaoyang and Colacelli, Ana and Lee, Matt and Xie, Le},
  journal={arXiv preprint arXiv:2509.07218},
  year={2025}
}

@article{azizi2026strengthening,
  title={Strengthening data center operations using grid-forming battery energy storage as a line-interactive uninterruptible power supply},
  author={Azizi, Arian and Morovati, Samaneh and Zamani, Amin and Piruzza, Joshua and Guo, David and Liu, Zhuoning and Taimela, Pasi},
  journal={International Journal of Electrical Power \& Energy Systems},
  volume={175},
  pages={111638},
  year={2026},
  publisher={Elsevier}
}

@article{pandey2026state,
  title={State-Space Model-Based Digital Twin With Real-Time Parameter Optimization Using MDA-EFO for Grid-Connected PV Inverter Systems},
  author={Pandey, Rituvic and Kumar, Nishant},
  journal={IEEE Transactions on Circuits and Systems I: Regular Papers},
  year={2026},
  publisher={IEEE}
}

@inproceedings{liu2026digital,
  title={Digital Twin-Enabled Modeling and Operational Optimization for Power-Electronics-Rich Distribution Networks},
  author={Liu, Zhiyang and Liu, Baoxia and Yu, Lichuan and Li, Yongxin and Li, Yuanhui},
  booktitle={2026 2nd International Conference on Electrical Automation and Artificial Intelligence (ICEAAI)},
  pages={605--610},
  year={2026},
  organization={IEEE}
}

@article{wu2025overview,
  title={An overview of digital twin technology for power electronics: State-of-the-art and future trends},
  author={Wu, Chenhao and Cui, Zhexin and Xia, Qian and Yue, Jiguang and Lyu, Feng},
  journal={IEEE Transactions on Power Electronics},
  year={2025},
  publisher={IEEE}
}

@article{jadhav2026digital,
  title={Digital twin-based cooling system optimization for data center},
  author={Jadhav, Shrenik and Liu, Zheng},
  journal={arXiv preprint arXiv:2603.01198},
  year={2026}
}

@article{sado2025digital,
  title={A digital twin based forecasting framework for power flow management in DC microgrids},
  author={Sado, Kerry and Peskar, Jarrett and Downey, Austin and Khan, Jamil and Booth, Kristen},
  journal={Scientific Reports},
  volume={15},
  number={1},
  pages={6430},
  year={2025},
  publisher={Nature Publishing Group UK London}
}

@inproceedings{nicolae2026opendt,
  title={OpenDT: Exploring Datacenter Performance and Sustainability with a Self-Calibrating Digital Twin},
  author={Nicolae, Radu and van der Toorn, Jules and Kraniti, Stavriana and Liu, Houcen and Iosup, Alexandru},
  booktitle={Companion of the 17th ACM/SPEC International Conference on Performance Engineering},
  pages={142--147},
  year={2026}
}

@manual{etran,
  author       = {{Electranix Corporation}},
  title        = {{E-TRAN} User Manual},
  organization = {Electranix Corporation},
  address      = {Winnipeg, MB, Canada},
  year         = {2024},
  note         = {Accessed: 2026-06-16},
  url          = {https://www.electranix.com/publication/e-tran-manual/}
}

@inproceedings{brown2020gpt3,
  author    = {Brown, T. B. and Mann, B. and Ryder, N. and Subbiah, M. and Kaplan, J. and others},
  title     = {Language Models are Few-Shot Learners},
  booktitle = {Advances in Neural Information Processing Systems (NeurIPS)},
  year      = {2020}
}

@article{huang2022PINN,
  title={Applications of physics-informed neural networks in power systems-a review},
  author={Huang, Bin and Wang, Jianhui},
  journal={IEEE Transactions on Power Systems},
  volume={38},
  number={1},
  pages={572--588},
  year={2022},
  publisher={IEEE}
}

@article{cao2020scalable,
  title={Scalable distribution systems state estimation using long short-term memory networks as surrogates},
  author={Cao, Zhiyuan and Wang, Yubo and Chu, Chi-Cheng and Gadh, Rajit},
  journal={IEEE Access},
  volume={8},
  pages={23359--23368},
  year={2020},
  publisher={IEEE}
}

@article{liu2015armax,
  title={ARMAX-based transfer function model identification using wide-area measurement for adaptive and coordinated damping control},
  author={Liu, Hesen and Zhu, Lin and Pan, Zhuohong and Bai, Feifei and Liu, Yong and Liu, Yilu and Patel, Mahendra and Farantatos, Evangelos and Bhatt, Navin},
  journal={IEEE Transactions on Smart Grid},
  volume={8},
  number={3},
  pages={1105--1115},
  year={2015},
  publisher={IEEE}
}

@article{rudin2021principles,
  title={Principles of mathematical analysis},
  author={Rudin, Walter},
  year={2021}
}

@article{garg2022can,
  title={What can transformers learn in-context? a case study of simple function classes},
  author={Garg, Shivam and Tsipras, Dimitris and Liang, Percy S and Valiant, Gregory},
  journal={Advances in neural information processing systems},
  volume={35},
  pages={30583--30598},
  year={2022}
}

@inproceedings{jena2026lilad,
  title={LILAD: Learning In-context Lyapunov-stable Adaptive Dynamics Models},
  author={Jena, Amit and Li, Na and Xie, Le},
  booktitle={Proceedings of the AAAI Conference on Artificial Intelligence},
  volume={40},
  number={26},
  pages={22164--22172},
  year={2026}
}

@article{jena2025llm,
  title={LLM-Based Adaptive Distribution Voltage Regulation Under Frequent Topology Changes: An In-Context MPC Framework},
  author={Jena, Amit and Ding, Fei and Wang, Jiyu and Yao, Yiyun and Xie, Le},
  journal={IEEE Transactions on Smart Grid},
  year={2025},
  publisher={IEEE}
}

@article{forgione2023system,
  title={From system models to class models: An in-context learning paradigm},
  author={Forgione, Marco and Pura, Filippo and Piga, Dario},
  journal={IEEE Control Systems Letters},
  volume={7},
  pages={3513--3518},
  year={2023},
  publisher={IEEE}
}

@article{radford2019language,
  title={Language Models are Unsupervised Multitask Learners},
  author={Radford, Alec},
  journal={OpenAI blog},
  year={2019}
}

\newpage

 





\end{document}